\documentclass[10pt,journal]{IEEEtran}

\usepackage{setspace}
\usepackage{cite,graphicx,amsmath,amssymb}
\usepackage{subfigure}
\usepackage{mdwmath}
\usepackage{mdwtab}
\usepackage{balance}
\usepackage{xcolor}
\usepackage{bm}
\usepackage{amsthm}
\usepackage{threeparttable}
\usepackage{algorithm}
\usepackage{algorithmic}
\usepackage{multirow}
\usepackage{flafter}
\usepackage{makecell}
\usepackage{diagbox}
\usepackage{graphicx,epstopdf}
\usepackage{epsfig}	

\graphicspath{{./src/img/}}

\newtheorem{theorem}{Theorem}

\newtheorem{lemma}{Lemma}

\newtheorem{corollary}{Corollary}

\allowdisplaybreaks
\title{Rethinking Integrated Sensing and Communication: When Near Field Meets Wideband}

\author{
        Zhaolin~Wang,  Xidong~Mu,  and Yuanwei~Liu
\thanks{The authors are with the School of Electronic Engineering
and Computer Science, Queen Mary University of London, London E1 4NS, U.K. (e-mail: zhaolin.wang@qmul.ac.uk, xidong.mu@qmul.ac.uk, yuanwei.liu@qmul.ac.uk).}
\vspace{-0.3cm}
}

\begin{document}

\maketitle

\begin{abstract}
This article revisits integrated sensing and communication (ISAC) systems that operate in the near-field region of large antenna arrays while utilizing large bandwidths. The article first describes the basic characteristics of a wideband sensing and communication (S\&C) channel, highlighting the key changes that occur during the transition from the far-field to the near-field region, namely \emph{strong angular delay correlations} and \emph{non-uniform Doppler frequencies}. It is then revealed that the near-field effect can facilitate wideband-like S\&C functionality, leading to efficient signal multiplexing and accurate distance sensing, and making large antenna arrays a viable alternative to large bandwidths. In addition, new capabilities for Doppler-domain signal multiplexing and velocity sensing enabled by non-uniform Doppler frequencies, which cannot be achieved by extending the bandwidth alone, are presented. Motivated by these results, several paradigm shifts required to leverage the full potential of near-field wideband ISAC systems are discussed, with particular emphasis on spectrum allocation, antenna array arrangement, transceiver architecture, and waveform design.

\end{abstract}

\begin{figure*}[t!]
    \centering
    \includegraphics[width=0.95\textwidth]{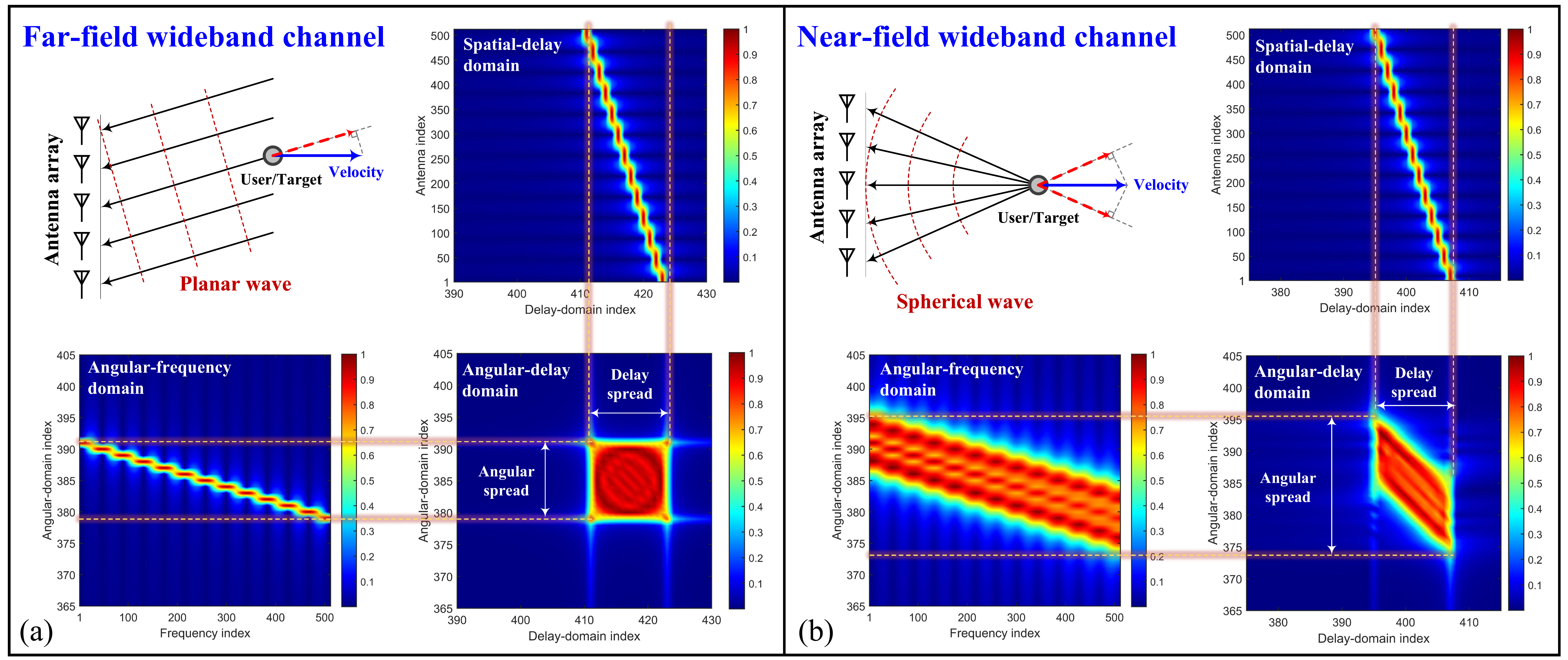}
    \caption{\textcolor{black}{Illustration of far-field ($200$ m) and near-field ($20$ m) wideband channels at $60$ GHz and direction $60^\circ$ with a $6$ GHz bandwidth, $512$ subcarriers, and $512$ antennas in spatial-delay ($\mathbf{H} \mathbf{F}_M^H$), angular-frequency ($\mathbf{F}_N^H \mathbf{H}$), and angular-delay ($\mathbf{F}_N^H \mathbf{H} \mathbf{F}_M^H$) domains, where $\mathbf{F}_N$ and $\mathbf{F}_M$ are normalized DFT matrices.}
    }
    \label{fig:channel}
\end{figure*}

\section{Introduction}
Integrated sensing and communication (ISAC) technique is emerging as a promising technique for future wireless networks, where the same radio frequency (RF) signals and hardware infrastructures is used for both sensing and communication (S\&C) purposes \cite{liu2022integrated}. ISAC technique is a key enabler for a variety of applications, such as automated driving, smart cities, and virtual reality. However, it also poses significant challenges in terms of waveform design, resource allocation, and performance evaluation. One of the key requirements of the ISAC technique is to achieve both high-resolution sensing and high-rate communication, which requires large bandwidth and large antenna arrays. Increasing bandwidth is one of the most straightforward methods to reach these objectives. On the one hand, the larger the bandwidth, the more data can be transmitted in a unit of time. On the other hand, large bandwidths also provide high temporal resolution and numerous samples for accurate ranging and Doppler frequency estimation. However, large bandwidths are not always available due to various factors such as regulatory restrictions, compatibility, standardization, and hardware limitations. 

Large antenna arrays have conventionally been considered to provide high spatial resolution in the angular-domain for tasks such as angle of arrival estimation and angular beamforming. Does this mean that large antenna arrays and large bandwidth are two independent paths for S\&C? No, as the apertures of antenna arrays reach extremely large sizes, coupled with extremely high carrier frequencies, the near-field effect becomes inevitable. The near-field and far-field are the two inherent electromagnetic field regions associated with antenna arrays. Due to the small aperture of the antenna array and the low carrier frequency, current wireless networks operate mainly in the far-field region. However, with the advent of Extremely Large Aperture Arrays (ELAA) and the use of higher frequency bands such as millimeter-wave and terahertz bands, recent studies have shown that the near-field region may extend to tens or even hundreds of meters \cite{10220205}. In certain situations, especially indoor scenarios, the near-field region may take precedence in S\&C applications. While the near-field effect increases the complexity of the S\&C channel, it can open up new opportunities to \emph{employ large antenna arrays as a substitute for large bandwidth} by enabling previously wideband-specific S\&C functions. The near-field effect also enables new S\&C functions that cannot be realized by extending bandwidth in high-mobility scenarios. The paradigm shift brought about by the near-field effect can be beneficial as well as detrimental, which requires a rethinking of ISAC system design. This is the main motivation of this article.

The main contributions of this article are summarized as follows:

\begin{itemize}
    \item Exploration of the fundamental characteristics of far-field and near-field wideband channels.
    \item A comprehensive comparison of the similarities and differences between near-field and wideband systems for S\&C, with an emphasis on how the near-field effect enables wideband-like functionalities and what fundamental changes it introduces in high-mobility scenarios.
    \item A rethinking of the evolution towards near-field wideband ISAC systems, with a specific focus on spectrum allocation, antenna array arrangement, transceiver architecture, and waveform design.
\end{itemize}

\section{Fundamentals of Wideband Channels: From Far-Field to Near-Field} \label{fundamental}

\textcolor{black}{In this section, we introduce the fundamentals of the far-field and near-field wideband channels with popular multi-carrier orthogonal frequency-division multiplexing (OFDM) modulation and fully digital antenna architecture \cite{sturm2011waveform} and highlight the changes introduced by the near-field effect. }
% For the sake of simplicity, we temporarily exclude the Doppler effect, which will be detailed in the subsequent sections.  

\subsection{Far-Field Wideband Channel} \label{fundamental_A}
We consider a wireless system comprising an $N$-antenna uniform linear array (ULA), $M$ equally-spaced subcarriers, and a moving communication user or sensing target. Let $\mathbf{H}[k] \in \mathbb{C}^{N \times M}$ represent the frequency-domain wideband channel matrix for the $k$-th OFDM symbol, where the entry $h_{nm}[k]$ represents the signal received at the $n$-th antenna and the $m$-th subcarrier. As shown in Fig. \ref{fig:channel}(a), in far-field systems, the signal phase changes can be approximated using the \emph{planar-wave model}. \textcolor{black}{Furthermore, user/target has the same velocity projection for all antennas, resulting in \emph{uniform Doppler frequencies}. Therefore, by omitting inter-carrier interference, $h_{nm}[k]$ can be modeled as \cite{liu2022integrated}
\begin{equation} \label{far-field_wideband}
    h_{nm}[k] = \beta e^{-j 2 \pi f_m (\tau - d_n \cos \theta) } e^{-j 2 \pi f_m k T_s v/c}.
\end{equation}    
Here, $\tau$, $\theta$, and $v$ denote the delay, angle of arrival, and velocity projection of the user/target relative to the antenna array center, respectively. Additionally, $\beta$ is the channel gain, $f_m$ is the subcarrier frequency, $d_n = (n-1 - \frac{N-1}{2}) d/c$ is the propagation time from the $n$-th antenna to the array center, $d$ is the antenna spacing, $c$ is the speed of light, and $T_s$ is the symbol duration.} In Fig. \ref{fig:channel}(a), the channel matrix $\mathbf{H}[k]$ is shown in different domains for a single symbol, obtained using discrete Fourier transform (DFT) \cite{wang2018spatial}. In the spatial-delay domain, wide bandwidth effectively resolves signal delay but large array aperture causes varied delays among antennas. In the angular-frequency domain, the large array resolves arrival angles but wide bandwidth links different frequencies to different angles. Therefore, there are notable delay and angular spreads in the angular-delay domain. The significant delay spread leads to reduced coherence bandwidth, known as the \emph{frequency-wideband effect}, causing potential inter-symbol interference. Meanwhile, the extensive angular spread, termed the \emph{spatial-wideband effect}, potentially causes beam squint or more severe beam split issues for the phased array.

\subsection{Near-Field Wideband Channel}
\textcolor{black}{As illustrated in Fig. \ref{fig:channel}, there are two key differences in near-field systems when compared to far-field systems: 1) the signal phase changes have to be modeled accurately using the \emph{spherical-wave model} and 2) the user/target has different velocity projections for different antennas, resulting in \emph{non-uniform Doppler frequencies}. Consequently, $h_{nm}[k]$ has to be modeled as \cite{10220205, wang2023near}:
\begin{equation} \label{near-field_wideband}
    h_{nm}[k] = \beta e^{-j 2 \pi f_m \sqrt{\tau^2 + d_n^2 - 2 \tau d_n \cos \theta }} e^{-j 2 \pi f_m k T_s v_n/c},
\end{equation} 
where $v_n$ denotes the velocity projection for the $n$-th antenna.} The resultant channel matrices for a single symbol in different domains are depicted in Fig. \ref{fig:channel}(b). \textcolor{black}{In the spatial-delay domain, the spherical wave propagation renders the delay of signals received at individual antennas from a linear to a non-linear function. However, this phenomenon is marginal unless the user/target is extremely close to the antenna array or has little shift from the broadside of the antenna array.} In the angular-frequency domain, angular spread occurs at individual frequencies. This is due to different antennas “seeing” the user/target from different directions in the near-field region, even within a single subcarrier. The near-field effect transforms the representation of channel matrices in the angular-delay domain from a rectangular to a diamond shape, where different delays correspond to different angular spreads, which implies a strong \emph{angular-delay correlation}. The angular spread in the angular domain is primarily caused by the intertwined angular and delay information. Such a correlation makes it possible to achieve wideband-like S\&C abilities within a single subcarrier. Additionally, the non-uniform Doppler frequencies can enable new functionalities that cannot be realized by wideband systems, which will be detailed in the following section.

\section{Near-Field versus Wideband: Similarities and Differences for S\&C} \label{difference}
In this section, we explore the wideband and near-field systems for S\&C. We note that while they have notable differences, they also share significant similarities. These shared characteristics pave the way for developing \emph{alternative near-field S\&C solutions to wideband S\&C solutions} and thereby position ELAAs as a potential alternative to the large bandwidth. Furthermore, the unique Doppler-frequency features in near-field systems can also \emph{facilitate new S\&C capabilities}. To better understand the similarities and differences between wideband and near-field S\&C systems, in the following, we discuss them from sensing and communication views, respectively, contrasting against the benchmark of far-field narrowband systems.

\begin{figure}[t!]
    \centering
    \includegraphics[width=0.45\textwidth]{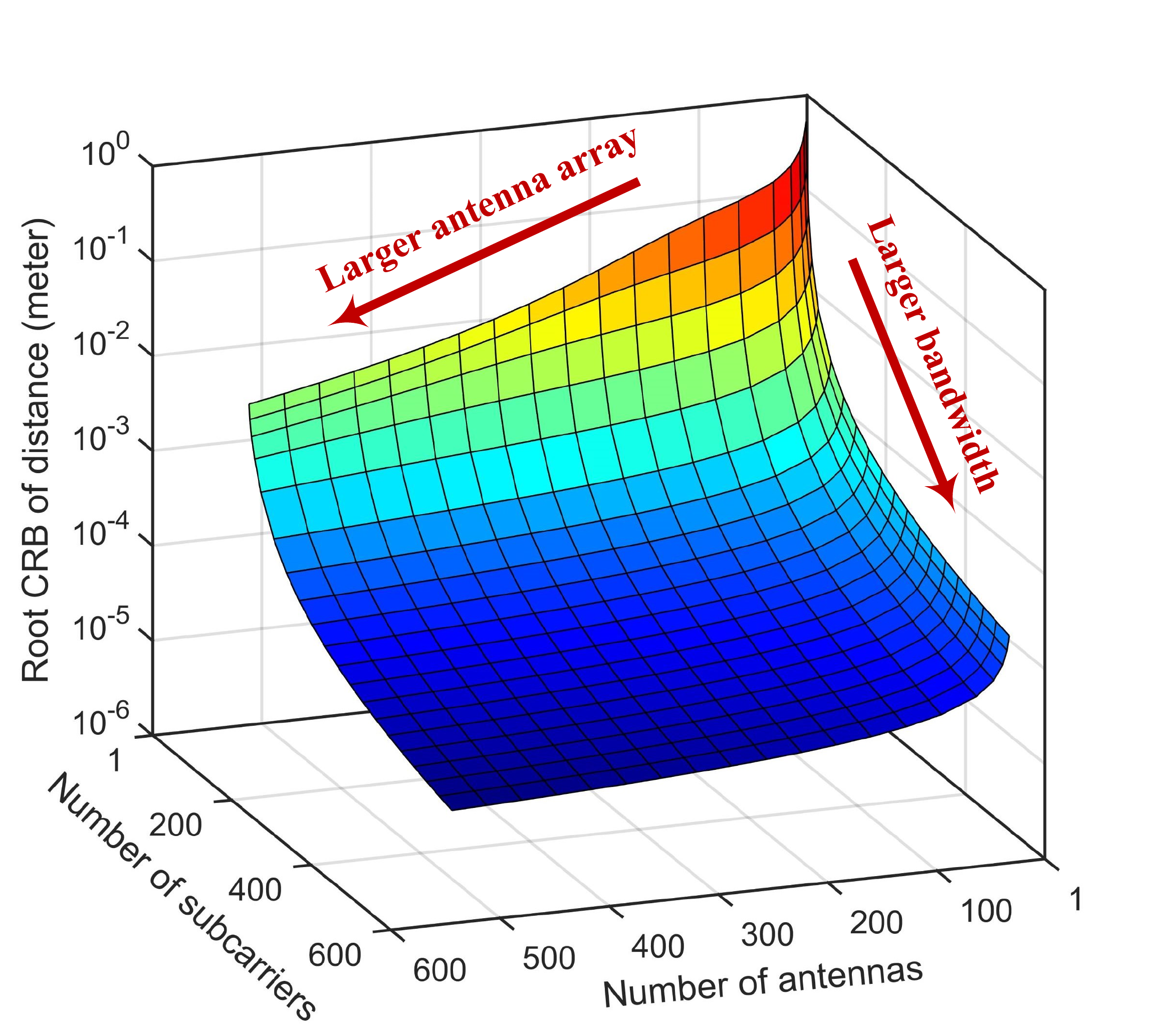}
    \caption{
    \textcolor{black}{The distance sensing CRB for a point-like target at $20$ m and direction $60^\circ$ achieved by a BS with $60$ GHz carrier frequency, half-wavelength antenna spacing, $1$ MHz subcarrier spacing, $1$ ms sensing duration, and $0$ dB SNR.}}
    \label{fig:CRB}
\end{figure}

\subsection{Sensing View}

Target parameter sensing typically entails estimating three key parameters: angle, distance, and velocity \cite{liu2022integrated}. Compared to far-field narrowband systems, where angle information can be effectively estimated through array processing, the advantage of wideband and near-field systems lies in the estimation of distance and velocity, which will be detailed in the following.

\subsubsection{Distance Sensing} 
In far-field narrowband systems, estimating distance is challenging for a single BS. In this case, locating an object requires the collaboration of at least two BSs, leading to high synchronization overhead. Wideband and near-field systems, however, provide different approaches that address this challenge effectively. 
\begin{itemize}
    \item \textbf{Wideband Solution:} In Section \ref{fundamental}, we discussed the reciprocal relationship between the delay domain and the frequency domain. Consequently, a wider bandwidth contributes to an enhanced resolution in the delay domain, facilitating the estimation of distance using a single sensing node. Certain wideband signals have been specifically designed for distance estimation, such as chirp signals, characterized by continuous frequency modulation \cite{mahafza2005radar}. Chirp signals enable distance estimation by measuring the time-of-flight through frequency shift. Additionally, many wideband communication signals are suitable for distance sensing. For example, leveraging OFDM signals allows for distance estimation through frequency-domain signal processing techniques \cite{sturm2011waveform}.
    
    \item \textbf{Near-Field Solution:} In near-field systems, the reliance on an extensive bandwidth for distance estimation is no longer necessary. Instead, distance estimation can be realized based on spatial-domain signal processing techniques, leveraging the spherical wave propagation of signals \cite{huang1991near}. This approach allows the near-field system to simultaneously estimate both the angle and distance of the target within the spatial domain. As a result, it becomes feasible to localize the target using a single sensing node and a limited bandwidth.
    
\end{itemize}

Comparatively, in near-field sensing, the accuracy of distance estimation diminishes as the distance from the target increases due to the weakening of the near-field effect, differing from the robustness of wideband sensing. This limitation indicates that near-field sensing cannot entirely replace the capabilities of wideband sensing. Hence, it is still beneficial to harness the wide bandwidth in near-field sensing to improve distance estimation. In Fig. \ref{fig:CRB}, we depict the performance of distance sensing in near-field wideband sensing systems, employing the Cramér-Rao bound (CRB) as the metric. The CRB sets the lowest limit for mean-squared error by an unbiased estimator under high signal-to-noise ratio (SNR) conditions. Notably, the CRB exhibits a decreasing trend with increasing bandwidth, attributed to the augmented delay resolution. Furthermore, a decrease in the CRB is observed with more antennas due to the amplified near-field effect. However, this decrease is less pronounced compared to the effect of a wider bandwidth. Nevertheless, near-field sensing demonstrates potential in achieving millimeter-level sensing accuracy within narrowband systems, effectively mitigating stringent bandwidth requirements.

\begin{figure}[t!]
    \centering
    \includegraphics[width=0.5\textwidth]{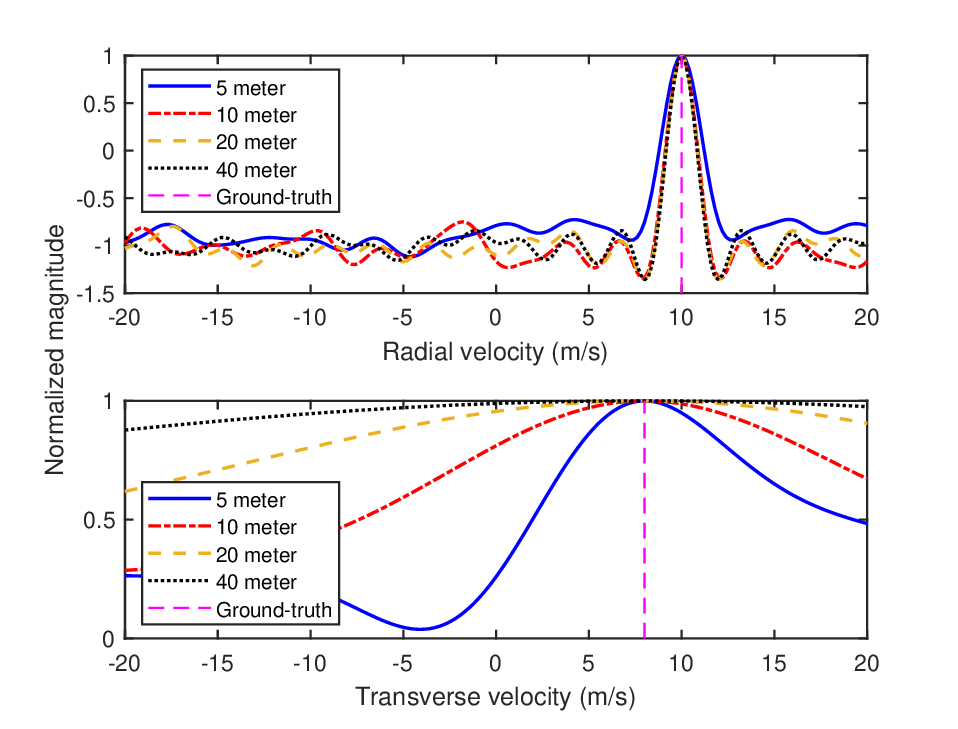}
    \caption{\textcolor{black}{Velocity profiles of single-target near-field sensing achieved by a BS with $28$ GHz carrier frequency, $512$ antennas, $100$ KHz bandwidth, and $2$ ms sensing duration.}}
    \label{fig:doppler}
\end{figure}

\subsubsection{Velocity Sensing}
The estimation of a target's velocity relies on detecting Doppler frequencies within signals. As we discussed in Section \ref{fundamental_A}, far-field systems exhibit uniform Doppler frequencies across all antennas introduced by the radial velocity, i.e., the velocity projection aligned with the direction between the BS and the target. In this case, since only partial velocity information is available, determining the target's motion status necessitates prior knowledge of the target's motion model \cite{liu2020radar} or collaboration among multiple BSs. 
Unfortunately, employing a large bandwidth does not modify the Doppler frequencies, rendering it ineffective in addressing this issue. Conversely, the near-field effect revolutionizes velocity sensing by introducing non-uniform Doppler frequencies that capture not only radial but also transverse velocities \cite{wang2023near}. This unique characteristic enables the estimation of the target's motion status, encompassing both the amplitude and direction of the velocity, using a single BS, without necessitating prior knowledge of the target's motion model. \textcolor{black}{An example of velocity profiles in narrowband near-field sensing is given in Fig. \ref{fig:doppler}, which exhibits discernible peaks around the ground-truth values for both radial and transverse velocities.} With full velocity information, predicting the target's subsequent location becomes simpler and more efficient. This heightened sensing capability can also greatly simplify various communication tasks like beam tracking, channel estimation, and cell handover that are sensitive to users' location. \textcolor{black}{Nonetheless, as shown in Fig. \ref{fig:doppler}, a larger distance reduces the dynamic range of transverse velocity sensing due to the diminishing near-field effect. This unique near-field ability underscores the importance of establishing a large near-field region to harness its advantages effectively.}

\begin{figure}[t!]
    \centering
    \includegraphics[width=0.5\textwidth]{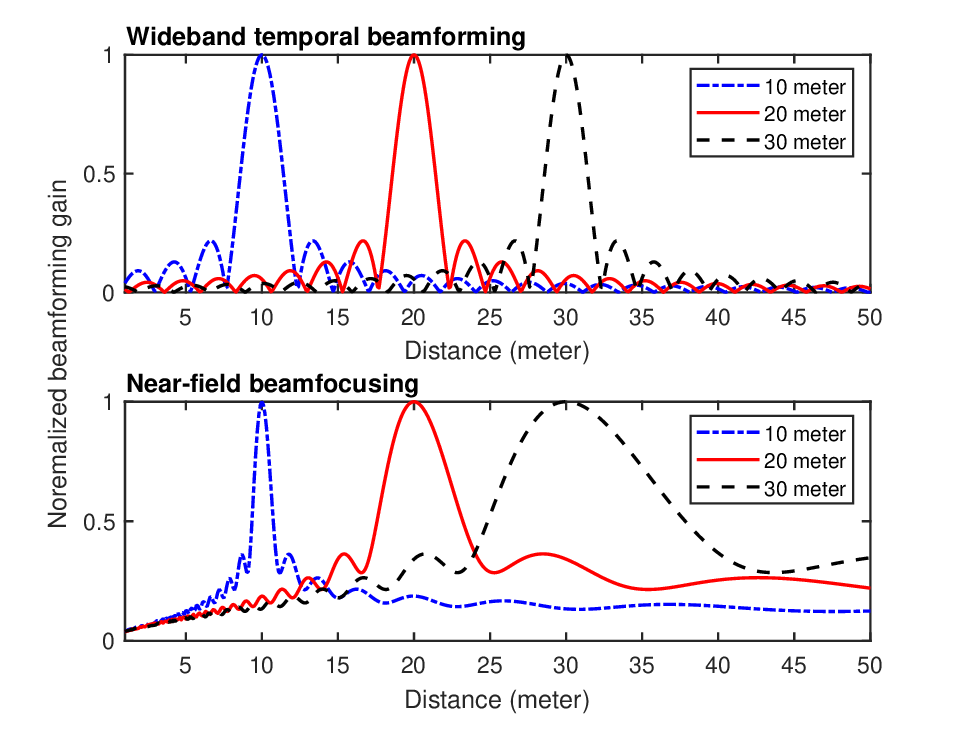}
    \caption{
        Comparison between wideband temporal beamforming ($1$ antenna and $512$ subcarriers with $0.5$ MHz spacing) and near-field beamfocusing ($1$ subcarrier and $512$ antennas with half-wavelength spacing.) realized by a BS with $28$ GHz carrier frequency. 
        }
    \label{fig:focusing}
\end{figure}

\begin{table*}[!t]
    \caption{\textcolor{black}{Comparison between Near-Field and Wideband S\&C Systems}}
    \label{table_1}
    \centering
    \resizebox{1\textwidth}{!}{
    \begin{tabular}{|c|l|l|l|}
    \hline 
     & \textbf{Operation} & \textbf{Wideband System} & \textbf{Near-Field System}  \\ \hline
    \multirow{2}{*}{\textbf{Sensing}} &Distance Sensing & \makecell[l]{Frequency-domain signal processing.}  & \makecell[l]{Spatial-domain signal processing.} \\ \cline{2-4}
    &Velocity Sensing &\makecell[l]{Estimation of only radial velocity. } &\makecell[l]{Estimation of both radial and transverse velocities.} \\ \hline
    \multirow{2}{*}{\textbf{Communication}} & Single-user signal multiplexing &\makecell[l]{Additional spectral DoFs} & \makecell[l]{Additional spatial DoFs} \\ \cline{2-4}
    &Multi-user signal multiplexing & \makecell[l]{Angular and temporal beamforming} & \makecell[l]{Beamfocusing and user-specific Doppler-frequency patterns}\\ \hline
    \end{tabular}
    }
\end{table*}

\subsection{Communication View}
Compared to far-field narrowband systems, both wideband and near-field systems offer significant improvements in communication data rates by multiplexing additional data streams, leveraging the frequency and spatial domains, respectively. In the subsequent discussion, we delve into their mechanisms for enhancing signal multiplexing by providing additional spectral and spatial degrees of freedom (DoFs), respectively, in both single-user and multi-user systems.

\subsubsection{Single-User Signal Multiplexing}
In narrowband single-user multiple-input multiple-output (MIMO) systems, the capacity for independent data streams relies on spatial DoFs. In far-field systems, the spatial DoFs are mainly determined by the uncorrelated paths between transceivers. However, at higher frequency bands characterized by significant scattering loss, the number of uncorrelated paths is limited, constraining the capacity for signal multiplexing.
\begin{itemize}
    \item \textbf{Wideband Solution:} Leveraging a wider bandwidth offers a direct solution to enhance signal multiplexing through additional spectral DoFs, where different data streams can be multiplexed on different subcarriers in the frequency domain.
    
    \item \textbf{Near-Field Solution:} The near-field system elevates signal multiplexing by increasing the spatial DoFs, i.e., the rank of MIMO channels in individual subcarriers, particularly associated with the line-of-sight (LoS) path. This arises from the receiver's capability to distinguish signals from different transmit antennas in both angular and distance domains. However, similar to near-field sensing, the spatial-domain multiplexing ability enabled near-field communication diminishes as the distance between transceivers increases due to the reduced spatial DoFs \cite{10220205}.

\end{itemize}

\subsubsection{Multi-User Signal Multiplexing}
Multi-user signal multiplexing is important for massive connectivity. In contrast to single-user systems, inter-user interference is another key factor that has to be considered in multi-user systems. In far-field systems, space-division multiple access can effectively distinguish users via angular beamsteering. However, a significant challenge arises when users are located in similar directions, causing significant inter-user interference.

\begin{itemize}
    \item \textbf{Wideband Solution:} Wideband systems can address this issue by allocating users located in similar directions to different subcarriers. However, this approach may underutilize available bandwidth resources as each subcarrier is only occupied by a subset of users. Alternatively, \emph{temporal beamforming} allows for sharing the entire bandwidth among all users and distinguishing users in the distance domain by using frequency subcarriers as a linearly deployed virtual antenna array \cite{wang2018spatial}. As depicted in Fig. \ref{fig:focusing}, with this virtual antenna array, temporal beamforming facilitates the generation of a signal directed towards a specific delay, enabling distance beamforming.

    \item \textbf{Near-Field Solution:} The angular-delay correlation in near-field systems enables distance beamforming in the spatial domain, which is referred to as the near-field \emph{beamfocusing} \cite{10220205}. As shown in Fig. \ref{fig:focusing}, unlike wideband temporal beamforming, beamfocusing is capable of distinguishing users at different distances using a single frequency subcarrier. However, beamfocusing exhibits a gradually reduced performance with increasing distance, whereas temporal beamforming maintains a consistent performance level. Apart from beamfocusing, it is also possible to distinguish users using the non-uniform Doppler frequencies in near-field systems. Consider a scenario involving two closely situated mobile users that cannot be distinguished in either angular or distance domains. If these users have different velocities, i.e., different magnitudes or directions, they will generate different non-uniform Doppler frequency patterns, which can be exploited to multiplex their signals. But, additional research efforts are required to develop corresponding Doppler-domain signal processing techniques.

\end{itemize}

Based on the above discussion, combining wideband and near-field systems can offer additional advantages. In single-user signal multiplexing, each subcarrier can accommodate more data streams through spatial-domain multiplexing enhanced by the near-field effect. In multi-user signal multiplexing, it is possible to simultaneously leverage the capabilities of near-field beamfocusing and wideband temporal beamforming to distance beamforming. Alternatively, the near-field beamfocusing and user-specific Doppler frequency patterns allow more users to be served by each subcarrier, thus enhancing spectral efficiency.

\begin{figure*}[t!]
    \centering
    \includegraphics[width=0.95\textwidth]{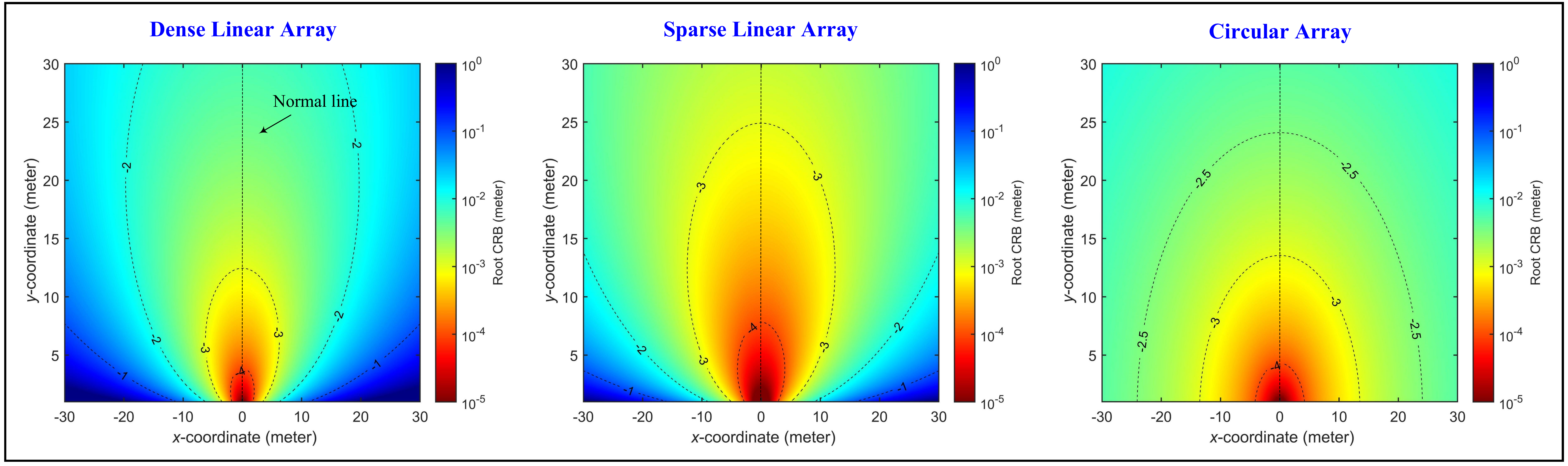}
    \caption{The distance sensing CRB for a point-like target with different array arrangements. Here, we consider a BS with $512$ antennas, $60$ GHz carrier frequency, and a single subcarrier. The antenna spacing of the sparse and dense linear arrays is set to one wavelength and half wavelength, respectively. The aperture of the circular array is equivalent to that of the dense linear array.
    }
    \label{CRB_map}
\end{figure*}

\section{Near-Field Wideband ISAC: The Way Forward}
Based on the discussion in previous sections, near-field systems can, to a certain extent, serve as a viable alternative to wideband systems and introduce new capabilities for S\&C. These advantages introduced by the near-field effect provide new possibilities for ISAC design. However, it may appear that developing new techniques that effectively coordinate near-field and wideband effects for enhancing ISAC performance is still in its early stages. In this section, we will revisit some recent research directions, with a particular focus on spectrum allocation, antenna array arrangement, transceiver architecture, and waveform design, and explore potential paths toward near-field wideband ISAC systems.

\subsection{Spectrum Allocation}
\textcolor{black}{In ISAC systems, the design of S\&C sub-systems can be based on either overlapped or non-overlapped spectrums. Far-field ISAC systems require spectrum sharing between S\&C due to certain functions, like distance sensing and massive connectivity, demanding a large bandwidth. Consequently, joint signal design across numerous subcarriers becomes crucial for a flexible and scalable S\&C tradeoff.
However, we note that S\&C functions may have different requirements for signal design \cite{xiong2023fundamental}. For example, in sensing, wide beams and deterministic signals may be preferred to cover a broad range and ensure stable sensing performance. In contrast, for communication, narrow beams and random signals may be favored to maximize capacity. While joint signal design shows potential for achieving high performance upper bound, the inherent diversity in S\&C requirements may lead to the very high complexity of optimizing the ISAC signals, potentially impeding real-time applications.}

\textcolor{black}{\emph{Rethinking: } In near-field wideband ISAC systems, \emph{the necessity for spectrum sharing can be substantially alleviated}, because of the strong spatial-domain signal processing capabilities enabled by the near-field effect. Consequently, S\&C sub-systems operating on non-overlapping frequency subcarriers may sufficiently meet essential S\&C requirements, which allows for optimizing S\&C signals independently for different objectives and reducing computational complexity. Moreover, to strike a balance between performance and complexity, and to address the S\&C trade-off, a partial spectrum-sharing strategy can be used, which partitions the entire bandwidth into three segments: a sensing-only band, a communication-only band, and an S\&C band. Adjusting the proportion of these bands allows for a flexible optimization of the performance-complexity trade-off and the S\&C trade-off.}

\subsection{Antenna Array Arrangement}
Current literature on ISAC commonly assumes that the antenna separation in an array is on the order of the sub-wavelength. This practice serves two primary purposes. On the one hand, according to the Shannon-Nyquist sampling theorem, a larger antenna spacing than half wavelength results in undesired grating lobes because of the spatial aliasing effect. On the other hand, the deployment of numerous densely spaced antenna elements contributes to enhanced spatial resolution in ISAC applications. Consequently, recent research efforts have concentrated on achieving spatially continuous transmitting and receiving surfaces.

\emph{Rethinking:} In near-field wideband ISAC systems, \emph{sparse antenna array becomes promising and considerations about the array shape become pivotal.} We note that the characteristics of both the near-field effect and the wideband effect closely relate to the aperture of the antenna arrays, rather than the number of antennas. In particular, a larger antenna array aperture boosts the near-field effect, facilitating better utilization of its unique advantages in ISAC systems. 
Installing more antennas is a straightforward approach to expand the array aperture, but it can escalate costs significantly, especially for fully digital antenna arrays where each element connects to a dedicated RF chain. To counter this, employing a sparse antenna array with larger spacing is more cost-effective, which can enlarge the aperture without additional antennas. Apart from increasing antenna spacing, optimizing array shapes is also a promising approach to increasing the effective array aperture. Linear and planar antenna arrays are widely employed in current wireless systems. However, these conventional array shapes exhibit reduced effective apertures when the user or target deviates from the array broadside. As a remedy, the circular or cylindrical array can be exploited \cite{10243590}, whose rotationally symmetric geometry ensures a consistent effective aperture for all directions. 
In Fig. \ref{CRB_map}, we present a numerical study on the distance sensing CRB, comparing three array arrangements: a dense linear array with half-wavelength spacing, a sparse linear array, and a circular array. As can be observed, the sensing performance of the dense linear array diminishes as the angle deviates from the normal line due to a reduced effective array aperture. The sparse linear array, on the other hand, significantly improves sensing performance across the entire region by increasing the array aperture but still exhibits significant performance loss near the end-fire region that has larger angles from the normal line. On the contrary, the circular array offers consistent sensing performance in all directions but demands more deployment space compared to linear arrays.

However, the array aperture cannot be increased without bound due to not only the physical limitation but also the potential adverse impacts. More specifically, a larger aperture also increases propagation delay differences at different antennas, amplifying both frequency-wideband and spatial-wideband effects. Mitigating the adverse impacts arising from these wideband effects can lead to additional signaling and hardware overhead \cite{gao2021wideband}. Based on the above discussion, the array arrangement design in near-field wideband ISAC systems should account for multiple factors, including grating lobes, frequency/spatial-wideband effects, and effective apertures. These factors may vary over time, particularly in scenarios involving mobile communication users and sensing targets. The recently emerged fluid antenna technique \cite{wong2020fluid} holds the promise of dynamically adjusting array arrangements in real-time to enhance near-field wideband ISAC performance.

\subsection{Transceiver Architecture}
In near-field wideband ISAC systems, the transition to higher frequency bands and larger bandwidths poses challenges in terms of power amplifier efficiency and analog-to-digital converter (ADC) power consumption, which are exacerbated when dealing with ELAAs. A popular approach to address these challenges is the use of hybrid analog and digital antenna array architectures, in which numerous antennas are connected to a limited number of RF chains via low-cost, energy-efficient analog phase shifters. In wideband systems, these phase shifters can be replaced with true-time delay units to facilitate frequency-dependent transmission \cite{gao2021wideband}.

\emph{Rethinking:} 
In near-field wideband ISAC systems, \emph{the hybrid analog and digital antenna array architecture faces may not be effective.} The reasons can be summarized in four-fold. \emph{Firstly}, in hybrid architectures, the number of effective spatial beams in individual subcarriers is limited by the number of available RF chains, which creates a dilemma for near-field beamfocusing. In particular, beamfocusing aims to employ different spatial beams to serve different users in angular and distance domains. However, the limited number of beams available in hybrid architectures requires the reuse of a single beam to serve multiple users, which makes beam focusing a hindrance rather than an advantage for signal multiplexing \emph{Secondly}, the enhancement of distance and velocity sensing capabilities in near-field systems depends on signals received by individual antennas with different angles and Doppler frequencies. However, hybrid architectures present an obstacle in acquiring these signals individually, as they are mixed in a shared RF chain. \emph{Thirdly}, from the resource allocation perspective, the near-field effect facilitates the low-complexity independent S\&C signal design on non-overlapping subcarriers. However, the hybrid architecture might impede this due to the need for jointly designing coefficients of the analog devices for all subcarriers. \textit{Lastly}, hybrid architectures may also hinder the dynamic adjustment of the array arrangement because the antennas are interconnected through complex analog networks. The above discussion suggests that embracing fully digital antenna arrays, in which each antenna is connected to a dedicated RF chain, may be critical to fully leverage the potential of near-field wideband ISAC systems. In this case, a low-resolution ADC can be employed to reduce cost and power consumption\cite{8310639}. By reducing the resolution, the power consumption of the ADC can be reduced exponentially. However, the impact of employing low-resolution ADCs on the performance of near-field wideband ISAC systems remains an open question.

\subsection{Waveform Design}

The waveform choice of wideband ISAC systems is a fundamental but certainly not finalized hot topic. Beyond the commonly used OFDM waveform explored in this article, several promising alternatives have been recently proposed, such as the orthogonal time frequency space (OTFS) waveform in the delay-Doppler domain \cite{9724198} and the single-carrier delay alignment modulation (DAM) waveform in the time-domain \cite{10105893}. These new waveforms often present appealing advantages compared to the traditional OFDM waveform, such as lower peak-to-average-power ratio, reduced guard interval, and increased tolerance for high Doppler frequencies.

\emph{Rethinking:} In near-field wideband ISAC systems, \emph{new waveforms may be needed to exploit the full potential of angular-delay correlation and non-uniform Doppler frequencies.} Most existing ISAC waveforms are designed by assuming the orthogonality between different domains. For example, in MIMO-OTFS systems, the angles, distances, and velocities of users/targets are independently handled in spatial, delay, and Doppler domains, respectively. However, the near-field effect breaks the orthogonality between these domains. Specifically, the angular-delay correlation makes it difficult to estimate angles and distances independently, and the non-uniform Doppler frequencies make the velocity information highly correlated with the spatial domain. \textcolor{black}{Additionally, the single-carrier DAM waveform can make the flexible spectrum allocation between S\&C challenging. Therefore, these existing waveforms may not be able to reach the full potential of near-field wideband ISAC systems. How to address the high correlation between different domains and enable flexible spectrum allocation in near-field wideband ISAC systems are challenging but pivotal issues for the next-generation ISAC waveform design.}

\section{Conclusion}

This article elucidated the fundamentals, challenges, and prospects of near-field wideband ISAC systems. The near-field effect empowers wideband-like S\&C capabilities within limited bandwidth and facilitates novel S\&C functionalities rooted in non-uniform Doppler frequencies. However, realizing the full potential of near-field wideband systems hinges on overcoming practical obstacles encompassing spectrum allocation, antenna array configuration, transceiver architecture, and waveform design, requiring further research efforts.

\bibliographystyle{IEEEtran}
\bibliography{reference/mybib}

\begin{IEEEbiographynophoto} {Zhaolin Wang} is a Ph.D. student at Queen Mary University of London, UK. He is the recipient of the Best Student Paper Award in IEEE VTC2022-Fall and the 2023 IEEE Daniel E. Noble Fellowship Award.
\end{IEEEbiographynophoto}

\begin{IEEEbiographynophoto} {Xidong Mu} (Member, IEEE) is currently a Postdoctoral Researcher at Queen Mary University of London, U.K. He received the IEEE ComSoc Outstanding Young Researcher Award for EMEA region in 2023. He is the recipient of the Best Paper Award in ISWCS 2022, the 2022 IEEE SPCC-TC Best Paper Award, and the Best Student Paper Award in IEEE VTC2022-Fall. 
\end{IEEEbiographynophoto}

\begin{IEEEbiographynophoto} {Yuanwei Liu} (Fellow, IEEE) is a senior lecturer at Queen Mary University of London, London, U.K. His research interests include NOMA, RIS/STAR, Integrated Sensing and Communications, Near-Field Communications, and machine learning. He serves as a Co-Editor-in-Chief of IEEE ComSoc TC Newsletter, an Area Editor of IEEE CL, and an Editor of the IEEE COMST/TWC/TVT/TNSE.
\end{IEEEbiographynophoto}

\end{document}